\documentclass[twocolumn,prl,amsmath,amssymb,showpacs]{revtex4}

\usepackage{bm}% bold math

\ifx\pdftexversion\undefined
  \usepackage[dvips]{graphicx}
\else
  \usepackage[pdftex]{graphicx}
\fi

\usepackage{color}

\def \be{\begin{equation}}
\def \ee{\end{equation}}
\def \bmlett{\begin{mathletters}}
\def \emlett{\end{mathletters}}

\def \ve{\varepsilon}

\def \pd{\phantom{\dagger}}

\def \ra{\rightarrow}

\def \hc{\hat{c}}
\def \hd{\hat{d}}
\def \hI{\hat{I}}

\begin{document}%******************************************************

%\bibliographystyle{simpl1}

%\onecolumn

\title{Shot Noise in SU(N) Quantum Dot Kondo Effects}
\author{P. Vitushinsky$^1$}
\author{A. A. Clerk$^1$}
\author{K. Le Hur$^2$}
\affiliation{$^1$Department of Physics, McGill University, Montr\'{e}al,
 Qu\'{e}bec, Canada, H3A 2T8}
\affiliation{$^2$Department of Physics, Yale University, New Haven,
 Connecticut,  USA, 06520}

\date{Sept. 14, 2007}

\begin{abstract}

We study shot noise in the current of quantum dots whose
low-energy behaviour corresponds to an SU($N$) Kondo model,
focusing on the case $N=4$ relevant to carbon nanotube dots.  For general $N$, two-particle Fermi liquid interactions
have two distinct effects:  they can enhance the noise 
via back-scattering processes with an $N$-dependent effective charge, 
and can also modify the coherent partition noise already present without interactions.
For $N=4$,  in contrast to the SU(2) case,  interactions enhance shot noise solely 
through an enhancement of partition noise.  This leads to a non-trivial prediction for experiment.
\end{abstract}

\pacs{72.70.+m, 72.15.Qm,  71.10.Ay, 73.63.Fg, 73.63.Kv}
% Kondo effect, Noise, Fermi Liquid Theory, Nanotubes, Quantum Dots}

\maketitle

Despite being more than 40 years old, the Kondo effect remains
among the most studied effects in condensed matter physics.  It is
a canonical example of how strong interparticle interactions, in a
relatively simple setting, can lead to highly non-trivial
behaviour; it is also particular attractive as it may be studied
in a controlled fashion in mesoscopic quantum dot
systems \cite{Glazman05}.
In spite of ultimately being the result of strong interactions,
much of the low-temperature phenomenology of the standard SU(2)
Kondo effect in quantum dots may be understood in terms of an
effective non-interacting model, having a resonant level (the
Kondo resonance) sitting at the Fermi energy.  This, however, is not the full story: 
low energy
properties are properly described by a Fermi liquid theory having
weak interparticle interactions \cite{Nozieres74, Nozieres78}.
While the effects of these interactions in quantum dot systems can be
subtle, it was recently shown by Sela et al.~\cite{Sela06}
that they play a large role in
determining the size of current fluctuations.  In particular,
Fermi liquid interactions lead to two particle scattering, and a
resulting universal effective charge $e^*$ (defined via the
magnitude of the backscattering current noise) that is larger than one: $e^* =
(5/3)e$.  This result represents one of the first clear
manifestations of the interacting nature of the low temperature
Kondo state in a quantum dot. 
Note that the fact $e^* \neq e$ only signals the contributions of two-particle
scattering to the current noise;
in contrast to fractional quantum Hall systems \cite{Glattli97},
%\nocite{Heiblum97}
it does not indicate the existence of fractionally charged quasiparticles.

In this paper, we now turn to slightly more exotic quantum dot
systems that are realizations of the higher symmetry SU($N$) Kondo
model ($N \geq 2$), and which are described by a corresponding
SU($N$)-symmetric Fermi liquid fixed point.  Such Kondo effects
involve the screening of an effective pseudospin that incorporates
both orbital and spin degrees of freedom; various physical
realizations of the $N=4$ case involving quantum dots have been
proposed \cite{Borda03,LeHur07,Choi05}
%\nocite{Zarand03, LeHur03, LeHur04,Lopez05}\nocite{Choi06}  
and even
measured  in carbon nanotube \cite{Jarillo05} 
%\nocite{Finkelstein07}
and vertical \cite{Sasaki04} dots.  
SU($N$) Kondo effects are 
interesting, as the ground state involves a non-trivial
entanglement of charge and orbital degrees of freedom. 

Here, we will examine how Fermi liquid
interactions in these systems affect the
small-voltage current noise $S_I$ through the quantum dot; note that
current noise in carbon nanotubes is already under
experimental study \cite{Yamamoto07}. 
Formulating the Fermi liquid theory in terms of scattering states 
and using the Keldysh technique to
calculate the non-equilibrium noise, we find that the situation is
markedly different from the SU(2) case.  For general $N$, interactions have
{\it two} distinct effects on shot noise.  The first (which is the only effect for $N=2$) is that 
Fermi liquid interactions lead to
new two-particle scattering processes; %at finite-voltage% 
these {\it enhance} the shot noise in a Poisonnian fashion (i.e.~via
uncorrelated tunneling events).  Such two-particle scattering can
result in the back-scattering of either one or two particles; we
denote the respective rates as $\Gamma_{1}$ and $\Gamma_{2}$.  We
find that the effective charges associated with these processes
are functions of $N$:
\begin{subequations}
\begin{eqnarray}
    e^*_1/e   & = & 2 T_0 -1 = 2  \sin^2( \pi/N)-1 \\
    e^*_2 & = & 2 e^*_1.
\end{eqnarray}
\label{eqs:estar}
\end{subequations}
Here, $T_0$ is the transmission coefficient through the dot at
zero temperature and voltage.  Eqs.~(\ref{eqs:estar}) reduce to the
results of Ref.~\onlinecite{Sela06} in the SU(2) case, while in
the SU(4) case, {\it both these effective charges vanish}.  For $N>4$, $e^*_1$ and $e^*_2$ 
become negative; as we show, this simply marks the transition from backscattering
to forward scattering.

In addition to the scattering-induced enhancement of current
noise, a second effect of Fermi-liquid interactions is to modify
the partition noise already present without Fermi liquid corrections.  Recall that in a phase-coherent conductor with transmission coefficient $T_0$, the partitioning of  electrons into transmitted and reflected streams results in $S_I \propto T_0 (1-T_0)$.  We find that two-particle Fermi-liquid interactions enhance this contribution, in 
keeping with more phenomenological studies of electron-electron interactions on shot
noise \cite{Blanter00}.  Note this effect is absent in the SU(2) case, as $T_0=1$.  In the especially relevant case $N=4$, this partition noise modification is the {\it only} effect of interactions; this leads to a parameter-free prediction for the interaction enhancement of noise  (c.f.~Eq.~\ref{eq:RatioInteracting}) which should be
highly amenable to experiment.  

{\it Model: } Though our approach is more general, for
concreteness we consider the case of SU(4) Kondo physics in a 
carbon nanotube dot \cite{Choi05}.  The system consists of a small
nanotube island (i.e.~the dot) attached via tunnel junctions to
two large nanotube leads.  In addition to having the
usual spin degree of freedom, electrons in this system also have
an orbital degree of freedom $m$ which can take one of two values
($m= +,-$);  $m$ labels two degenerate bands in the carbon
nanotube \cite{Choi05}.  We further assume that $m$
is conserved in lead-dot tunneling; this appears to be consistent with
experiment \cite{Jarillo05}.   Due to strong on-site Coulomb interactions, 
the quantum dot can be tuned to a regime where
it is only singly occupied.  The dot then acts as an effective 4-component pseudospin,
characterized by both its value of physical spin and its orbital
index $m$.  Further, dot-lead tunneling becomes an effective
exchange-like interaction between the dot and conduction electron pseudospin.  The resulting model is the SU($4$) Kondo model.  Renormalization group arguments \cite{Choi05} show
that at low temperatures, its physics is described a strong
coupling fixed point.  The
ground state corresponds to a singlet  formed by the impurity
pseudospin (the quantum dot) and a screening cloud of electrons in
the leads. Note that this model could be easily generalized to an
SU($N$) ($N>4$) Kondo model by allowing the orbital index $m$ to
take on $N/2$ different values (i.e.~there are $N/2$
degenerate electron bands).  In what follows, we will consider the general $SU(N)$
case, paying special attention to the SU(4) case. We will use
$\sigma$ to denote one of the $N$ degenerate electron ``flavours".

{\it Fermi liquid theory using scattering states:} We will be
interested in the behaviour of our system at energies
much smaller than $T_K$, the Kondo temperature.  In this
regime, the SU($N$) Kondo effect is described by a local Fermi
liquid theory \cite{Nozieres78, Nozieres80}, which describes both
elastic scattering off the Kondo singlet, as well as interparticle
interactions induced by virtual excitations of the singlet.  Both
effects are described by the phase shift $\delta_{tot,\sigma} =
\delta_0 + \delta_\sigma(\ve)$ for s-wave scattering of an
electron with pseudospin $\sigma$ off the impurity.  $\delta_0$ is
the phase shift at the Fermi energy; its value is determined by
the Friedel sum rule to be $\delta_0 = \pi/N$.
$\delta_\sigma(\ve)$ accounts both for the energy-dependence of
elastic scattering off the effective impurity, as well as for
inelastic processes:
\begin{eqnarray}\label{eq: FL phase shift}
    \delta_{\sigma}(\ve)=
     \frac{\alpha \ve}{k_B T_K} +
    \alpha' \left( \frac{\ve}{k_B T_K} \right)^2  -
        \frac{\beta \sum_{\sigma' \neq \sigma} n_{\sigma'}}{\nu k_B
        T_K}\;.
        \label{eq:ElasticPhase}
\end{eqnarray}
Here, $\nu$ is the density of states per electron flavour and the
energy $\ve$ is measured from Fermi level.   Note that we retain the $\ve^2$ dependence of the elastic phase
shift; while it plays no role
in the SU(2) case, it will play a critical role for $N=4$.     Universality at the 
SU(N) Kondo fixed point
ensures that the ratios between $\alpha, \alpha'$ and $\beta$ are
fixed.  The pinning of the energy of the Kondo resonance 
%(which for $N>2$ is above the Fermi energy) 
relative to the Fermi energy leads to the relation
$\alpha = (N-1) \beta$ \cite{Nozieres80}.  For $N=2$, the fixed
point has particle-hole symmetry, implying $\alpha'=0$.  For $N>2$, 
the Kondo resonance is above the Fermi energy; consequently, the
fixed point is not particle-hole symmetric.  For $N=4$, one finds $\alpha' = \alpha^2$.
This follows from equating the elastic parts of the phase shift $\delta_{\sigma,tot}$
with $\arctan(\Gamma/(2 (\ve_d - \ve))$, the phase shift of a resonant level (i.e.~the Kondo resonance) having width $\Gamma$ and energy $\ve_d$.  For $N=4$, expanding to order $\ve$ yields $\ve_d = \Gamma/2$ and $\Gamma = T_K / \alpha$;  expanding to $\ve^2$ yields $\alpha^2 = \alpha'$.  These properties of the $N=4$ Kondo resonance are consistent with recent NRG calculations \cite{Choi05}.

Applying the Fermi liquid picture to our quantum dot system, we
may replace the dot plus two lead system by a single
one-dimensional channel described by a field operator
$\hat{\psi}_{\sigma}(x)$:  for $x < 0$, this describes the left
lead, while for $x > 0$, it described the right lead.  The
scattering (both elastic and inelastic) due to the quantum dot
plus screening cloud may then be modeled as occurring locally at
$x=0$.  It is convenient to work in a basis of scattering states which
correspond to the Fermi-energy phase shift $\delta_0$.  
This can be converted to a transmission probability
$T_0$ via the usual relation $T_0 = \sin^2 \delta_0$.
We thus introduce scattering states $\phi_L(x;k \sigma)$ and
$\phi_R(x; k \sigma)$ which correspond (respectively) to waves incident on
the dot from the left or from the right lead.  For simplicity, we focus throughout 
on the case where both leads are equally coupled to the dot; 
deviations will be considered in \cite{futurepaper}.
Inversion symmetry then implies that for $x < 0$:
\begin{eqnarray}\label{eq: L scattered state}
     \phi_L(x; k\sigma) & = &   e^{i (k_F+k) x} - \sqrt{1-T_0} e^{-i (k_F+ k) x} \\
     \label{eq: right scattered state}
     \phi_R(x; k\sigma) & = &  - i \sqrt{T_0} e^{-i (k_F+k) x}\;.
\end{eqnarray}

Letting $\hc_{L,k\sigma},\hc_{R,k\sigma}$ denote the anhiliation
operators for the scattering states, 
the field operator may be written as
$\hat{\psi}_\sigma(x) \propto \sum_ {k} \left(   \phi_L(x; k\sigma) \hc_{L,k\sigma} +
\phi_R(x; k\sigma) \hc_{R,k\sigma} \right)$.  The current operator for $x<0$ then takes the form:
\begin{eqnarray}
\label{eq:IDefn}
    \hI(x) & = &    \hI_{D}(x)+\hI_{OD}(x)\;,
\end{eqnarray}
with ($R_0 \equiv 1- T_0$)
\begin{subequations}
\begin{eqnarray}
%\begin{array}{lcl}
    \hI_{D}(x) &=& \frac{e}{h \nu} \sum_{k,k';\sigma}
    \Bigg[
        \hc^{\dag}_{R,k' \sigma} \hc^{\pd}_{R,k \sigma}
         [- T_0 e^{-i (k-k') x}]
         \nonumber \\
          &&+\hc^{\dag}_{L,k' \sigma} \hc^{\pd}_{L,k \sigma}
         [ e^{i (k-k') x } - R_0 e^{-i (k-k') x}]\Bigg]
	\label{eq:ID}
          \\
 \hI_{OD}(x) &=&  \frac{e}{h \nu} \sum_{k,k'; \sigma}
            \label{eq:IOD}
	\\
                 && \Bigg[
            \hc^{\dag}_{L,k' \sigma} \hc^{\pd}_{R,k \sigma}
             [- i \sqrt{T_0 R_0} e^{-i (k-k') x}]   +
             \textrm{h.c.} \Bigg]\;.
             \nonumber
%             \end{array}
\end{eqnarray}
\end{subequations}
Without Fermi liquid corrections, the average current would be determined by $\hI_{D}$, while the zero-temperature current noise would be determined by $\hI_{OD}$
\cite{Blanter00}.  

Similar to the $N=2$ case \cite{Glazman05, Affleck90}, the effects of $\delta_\sigma(\ve)$ may be incorporated via the Hamiltonian:
\begin{subequations}
\begin{eqnarray}
    H & \equiv &
        \sum_{k, \sigma,j=L,R} \xi_k \hc^{\dag}_{j,k \sigma} \hc^{\pd}_{j,k \sigma}
        + H_{\alpha} + H_{\alpha'} + H_{\beta} \\
        \label{eq:HDefn}
    H_{\alpha}  & = &
        - \frac{\alpha}{2 \pi \nu T_K} \sum_{k,k',\sigma}
            \left( \xi_k + \xi_k' \right) \hd^{\dag}_{k \sigma} \hd^{\pd}_{k' \sigma}
        \label{eq:HAlpha}  \\
    H_{\alpha'}  & = &
        - \frac{\alpha'}{2 \pi \nu (T_K)^2} \sum_{k,k',\sigma}
            \left( \xi_k + \xi_k' \right)^2 \hd^{\dag}_{k \sigma} \hd^{\pd}_{k' \sigma}
        \label{eq:HAlpha}  \\
    H_{\beta}  & = &
      \frac{\beta}{2 \pi \nu^2 T_K} \sum_{k',q',k,q}
        \sum_{\sigma \sigma'}
        \hd^{\dag}_{k\sigma} \hd^{\pd}_{k'\sigma}
        \hd^{\dag}_{q \sigma'} \hd^{\pd}_{q' \sigma'}
        \label{eq:HBeta}\;,
\end{eqnarray}
\end{subequations}
Here, $\xi_k = \hbar v_F k$ and $\hd_{k \sigma} = (\hc_{L,k \sigma} + \hc_{R,k \sigma}) /
\sqrt{2}$ (only the symmetric combination enters due to inversion symmetry).
$H_\alpha
+ H_{\alpha'}$ and $H_{\beta}$ represent, respectively, the elastic scattering
from the impurity (to order $(\ve / T_K)^2$) and the presence of two-particle scattering.  This Hamiltonian explicitly reproduces the phase shift $\delta_{\sigma}$
(Eq.~(\ref{eq: FL phase shift})); its form follows from the usual relation between scattering
and potential matrices (see, e.g., \cite{Clerk03}).
The scattering
description presented here is consistent with the
results obtained in Ref.~\onlinecite{LeHur07} for $N=4$ using
conformal field theory arguments.  As we will show, using this
Hamiltonian and the current operator in Eq.~(\ref{eq:IDefn})
allows one to both recover results for the current obtained 
from $\delta_{tot,\sigma}$, {\it as well as} calculate
current fluctuations.

{\it Effect of elastic scattering:} We focus on the regime of zero
temperature, and small voltage $\mu_L - \mu_R = eV \ll T_K$.
As we treat the case of equal couplings to the
left and right leads, we use a symmetric voltage bias: $\mu_L = -
\mu_R = eV/2$ (i.e.~the Kondo resonance position is fixed
relative to $(\mu_L + \mu_R)/2$).
Consider the first the effects of $H_{\alpha}+H_{\alpha'}$, which
describe purely elastic scattering.  Keeping only these terms, our
system is equivalent to a non-interacting system having
an energy-dependent transmission coefficient $T(\ve)$ given by:
\begin{eqnarray}
    T(\ve) = \sin^2 \left[\delta_0 + \alpha \frac{\ve}{k_B T_K}
    + \alpha' \left( \frac{\ve}{k_B T_K} \right)^2 \right]\;.
\end{eqnarray}
We have verified through an explicit perturbative calculation in 
$H_{\alpha}+H_{\alpha'}$ that to order $(eV/T_K)^3$, 
the elastic scattering contributions to
the current and noise do indeed correspond to $T(\ve)$.  In particular, the elastic contributions to the
average current are given by
\footnote{For an asymmetric bias, we find for $N=4$ a $V^2$ term in the current, 
exactly as found in Ref.~\onlinecite{LeHur07}.  In this case, $\alpha'$ plays no
role to leading order.}:
\begin{eqnarray}
    \langle I \rangle \Big|_{\beta=0} & \equiv &
        \frac{N e^2}{h} \int_{\mu_R}^{\mu_L} d \ve
            T(\ve) \\
    & = &
        \frac{Ne^2 V}{h} \Big[
            T_0 +
            \frac{\alpha^2 \cos2 \delta_0
            + \alpha' \sin 2 \delta_0}{12}
            \left( \frac{eV}{k_B T_K} \right)^2
            \nonumber \\
            &&
            + O((eV/k_B T_K)^4 ) \Big]
            \label{eq:IAlpha}\;.
\end{eqnarray}
Here, $T_0 = \sin^2 \delta_0$.  Note for $N=4$, there is no $\alpha^2$
contribution to the current, as $\delta=\pi/4$. 
%Note that second-order contribution
%to the elastic phase shift (proportional to $\alpha'$) never plays
%a role in the case $N=2$, as $\delta_0 = \pi/2$.  In contrast, it
%plays a crucial role for $N=4$, as in this case
%$\delta_0 = \pi/4$.

Similarly, the elastic scattering contributions to the
zero frequency current noise $S_I \equiv 2 \int dt \langle \delta
I(t) \delta I(0) \rangle$ are given by the non-interacting
formula:
\begin{eqnarray}
    S_I \Big|_{\beta=0}& \equiv &
        \frac{2 N e^3}{h} \int_{\mu_R}^{\mu_L} d \ve
            T(\ve) (1 - T(\ve) )
            \\
    & = &
        \frac{2 Ne^3 V}{h} \Big[
            T_0 (1-T_0)
             \label{eq:SIAlpha} \\
            &&
            +
                \frac{\alpha^2 \cos(4 \delta_0) + \alpha' \sin(4 \delta_0)/2}{12}
                \left( \frac{eV}{k_B T_K} \right)^2
             \nonumber \\
                        &&
            + O((eV/k_B T_K)^4 ) \Big] \nonumber
\;.
\end{eqnarray}
Thus, the elastic part of the Fermi liquid phase shift $\delta_\sigma(\ve)$ yields $V^3$ terms in both the current and current noise.  In the especially relevant case of $N=4$, their ratio is simply: 
\begin{eqnarray}
	\label{eq:NonintRatio}
   \frac{1}{2e} \frac{ d^3 S_I / dV^3} { d^3 \langle I \rangle / d V^3 } \Big|_{\beta=0,V=0}
    & = & - \frac{\alpha^2}{ \alpha'} = -1\;.
\end{eqnarray}
This is just the result for a non-interacting resonant-level.
As we will see, two-particle interactions modify both the sign and magnitude
of this ratio; this relative enhancement of noise over a non-interacting model 
is our key prediction for experiment.

{\it Effect of two-particle scattering:} We now turn to the
effects of two-particle scattering at the Fermi liquid fixed
point, described by $H_{\beta}$ (Eq.~(\ref{eq:HBeta})).  We
have calculated its effects on the average current and current
noise perturbatively using the Keldysh approach.
On a
heuristic level, $H_{\beta}$ induces two-particle scattering
between the scattering states $\phi_L$, $\phi_R$.  For $\mu_L =
\mu_R + eV$ ($eV >0$) and zero temperature, there is a range of
energies $\ve$ where the $\phi_L(\ve)$ states are occupied, while
the $\phi_R(\ve)$ are unoccupied. As a result, 
three types of inelastic processes are possible: (a)
$\phi_L\phi_L\to\phi_R\phi_L$, (b) $\phi_L\phi_R\to\phi_R\phi_R$,
(c) $\phi_L\phi_L\to\phi_R\phi_R$. Using Fermi's Golden Rule in
$H_\beta$, one finds that for a fixed initial and final spin configuration, 
processes (a) and (b) occur at a rate
$\Gamma_1$, while process (c) occurs at a rate $\Gamma_2$. These rates
have no explicit-$N$ dependence, and are given by 
$\Gamma_1 = \Gamma_2/8 = \beta^2\frac{1}{24}\, \frac{eV}{h}
\left(\frac{eV}{k_B T_K}\right)^2$, as in Ref.~\cite{Sela06}.

The current associated with these interaction-induced scattering
processes does not follow immediately from the rates, as unlike
the SU(2) case, $T_0 \neq 1$ and hence the scattering states
$\phi_L, \phi_R$ are not eigenstates of current.  A long but
straightforward calculation using the Keldysh technique and
treating $H_\beta$ as a perturbation yields the following
contribution to the current to order $\beta^2$:
\begin{eqnarray}
    \delta \langle I \rangle \Big|_{\beta} & = & -
               (N_a+N_b) e^*_1 \Gamma_1 - N_c e^*_2 \Gamma_2 
        \label{eq:IBeta}\;.
\end{eqnarray}
Here, the ``effective charges" $e^*_1, e^*_2$ are exactly the
charges given in Eq.~(\ref{eqs:estar}).  The factors $N_a, N_b,
N_c$ simply count the multiplicity of each of the three processes
(i.e.~the number of distinct initial and final states once
pseudospin is included). One finds $N_a = N_b = 2 N_c =  N(N-1)$.
The contribution to the current given in Eq.~(\ref{eq:IBeta}) can
be combined with that in Eq.~(\ref{eq:IAlpha}) to yield the total
low voltage current to order $(eV / T_K)^3$.

There is a simple heuristic way to understand the effective
charges appearing in Eq.~(\ref{eq:IBeta}):  they are proportional
to the {\it total change in the scattered wave amplitude
associated with a particular interaction-induced scattering
process}.  An interaction-induced scattering event leads to a
sudden change in the amplitude of the scattered waves associated
with a particular pair of scattering states.  For example,
consider the $\Gamma_2$ process, which takes a pair of $\phi_L$
scattering states to a pair of $\phi_R$ scattering states 
(i.e.~$\phi_L\phi_L \ra \phi_R\phi_R$); without loss of generality,
consider the current in the left lead. Initially, we have two left
scattering states; the associated scattered-wave amplitude (c.f.
Eq.~(\ref{eq: L scattered state})) is thus $2 \times (1-T_0)$. In
the final state, we have instead two right scattering states; the
associated scattered-wave amplitude (c.f. Eq.~(\ref{eq: right
scattered state})) is thus $2 \times T_0$. The $\Gamma_2$ process
thus suddenly changes the scattered wave amplitude by an amount $
2(2 T_0-1) = e^*_2/e$. A sudden change in scattered wave amplitude
implies a delta-function blip in the current of magnitude $e^*_2$:
$\delta I(t) \simeq e^*_2 \delta(t)$.  Such ``blips" occur at a
rate $\Gamma_2$, hence the expression in Eq.~(\ref{eq:IBeta}).  A
similar analysis may be given of the $\Gamma_1$ processes and the
charge $e^*_1$.  We stress that Eq.~(\ref{eq:IBeta}) follows from
a rigorous perturbative calculation of the current; the result
agrees in the SU(2) and SU(4) case with results obtained by
alternate methods \cite{Sela06, LeHur07}.

The utility of the interpretation of Eq.~(\ref{eq:IBeta}) in terms
of effective charges becomes apparent when we now turn to the
interaction contribution to the current noise $S_I$.  Based on the
effective charge interpretation, we would expect a Poissonian
contribution to the current noise:
\begin{eqnarray}
    \delta S_I  \Big|_{\beta,D} & = &
        2(N_a+N_b) \left( e^*_1 \right)^2 \Gamma_1 +
        2 N_c \left( e^*_2 \right)^2 \Gamma_2
        \label{eq:SIBeta1}\;.
\end{eqnarray}
This is {\it precisely} the answer we find in our full calculation when
we sum all contributions which are independent of $\hat{I}_{OD}$
(c.f Eq.~(\ref{eq:IOD})); in the case where $T_0 \ra 1$ or $T_0 \ra 0$,
these are the only contributions, as $\hI_{OD} = 0$.
Thus, we see that
interaction-induced scattering events enhance the shot noise in a
very simple way, namely via an uncorrelated Poisson process. For
$N=2$, these contributions reduce to what was found in
Ref.~\onlinecite{Sela06}.  {\it For $N=4$, the effective charges
$e^*_1$ and $e^*_2$ are zero, and these scattering processes make
no contribution to the noise!}

We now turn to the second interaction-induced contribution to the noise, namely
the contribution from diagrams involving $\hI_{OD}$ (c.f.~Eq.~(\ref{eq:IOD})).  Recall that 
this part of the current operator is responsible for the partition shot noise already
present without Fermi liquid corrections (i.e.~first term in Eq.~(\ref{eq:SIAlpha})).  We find that Fermi-liquid two-particle interactions {\it enhance} this noise; summing all diagrams involving $\hI_{OD}$ to order $\beta^2$, we have:
%  prefactor is 2 - (4/3) = + 2/3
\begin{eqnarray}
	&&    \delta S_I  \Big|_{\beta,OD}  =    
	      \label{eq:SIBeta2} \\
	&& \left( 2(N-1) -\frac{8}{3} \right)
        N(N-1) 
        \frac{e^3V}{h}T_0 R_0 \beta^2 \left(\frac{eV}{k_B
        T_K}\right)^2\;. \nonumber
\end{eqnarray}
As will be discussed in Ref.~\cite{futurepaper}, the dominant effect here
is an enhancement of noise due to an interaction-induced enhancement 
of the particle-hole density of states.  Note that in the $N=2$ case this contribution vanishes, as there is no partition noise in this case.  Also note that it is 
impossible to express this contribution purely in terms of the
scattering rates $\Gamma_1$ and $\Gamma_2$ introduced above.

Focusing now on the experimentally-relevant case $N=4$, we may again compare
the non-linear-in-$V$ dependencies of the current noise and current.  Combining all contributions (Eqs. (\ref{eq:IAlpha}, \ref{eq:SIAlpha}, 
\ref{eq:IBeta}, \ref{eq:SIBeta1}, \ref{eq:SIBeta2})), we find
\begin{eqnarray}\label{eq:IntRatio}
    \frac{1}{2e}\,\frac{ d^3 S_I / dV^3} { d^3 \langle I \rangle / d V^3 } \Big|_{V=0}
    & = &  \frac{-\alpha^2 +15 \beta^2}{ \alpha'} = \frac{2}{3}\;.
    \label{eq:RatioInteracting}
\end{eqnarray}
Comparison against Eq.~(\ref{eq:NonintRatio}) shows that this ratio is modified by a factor of $-2/3$ compared to the non-interacting expectation.  This universal interaction-induced noise enhancement is a central result of this Letter;
it is a stringent test of the theory (involving three Fermi-liquid parameters), and should
be amenable to experiment.  Although Eq.~(\ref{eq:IntRatio}) was derived for a dot symmetrically coupled to the leads,
it is largely insensitive to small deviations from this limit; the same is true with
the presence of residual potential scattering at the fixed point \cite{futurepaper}.
%It should also be noted that the ratio on the left-hand side of Eq.~(\ref{eq:IntRatio}) is the same in the SU($2$) case \cite{Sela06}; this correspondence however is completely coincidental.  In the SU($2$) case, Fermi-liquid interactions {\it increase} the noise via interaction-induced scattering processes (c.f.~Eq.(\ref{eq:SIBeta1})); in contrast, in the SU($4$) case the interaction contribution is {\it negative} and solely due to the suppression of partition noise.  
We stress again that in the SU(4) case, Fermi-liquid interactions enhance shot noise
through a modification of the coherent partition noise.  This is in contrast to the SU(2) case, where the modification is due to interaction-induced scattering events.
Note also that in general, the ratio in Eq.~(\ref{eq:IntRatio}) is a complex function of $N$.

In conclusion, we have shown how Fermi liquid interactions affect 
current noise through a SU($N$) Kondo quantum dot, identifying two distinct
physical mechanisms.  In the experimentally-relevant $N=4$ case, we show that two-particle interactions modify the shot noise in a non-trivial and universal manner; this serves as a robust prediction for experiment.  Our approach in this Letter is also well suited to study the effect of interactions on current noise in a variety of mesoscopic conductors \cite{Buttiker07}.  

We thank T. Kontos for useful discussions, and especially C. Mora for helping us find the error in our original calculation.  A.C. and K.L.H. acknowledge support from NSERC and CIFAR.

{\it Note Added}: After the initial submission of our paper, a manuscript by C. Mora et al.~appeared (arXiv:0709.2089), looking at SU(N) Kondo shot noise at finite temperature.
Our results agree with the corrected version of this manuscript at T=0.
\bibliographystyle{apsrev}
\bibliography{su4refs}

\end{document}